\documentclass[review]{elsarticle}    

\usepackage{fullpage}
\usepackage{graphicx}
\usepackage{amsmath}
\usepackage{amssymb}
\usepackage{multicol}
\usepackage{multirow}
\usepackage{caption}
\usepackage{subcaption}
\usepackage{amsthm}

\journal{Operations Research for Health Care}
\begin{document}
\begin{frontmatter}

\title{Proactive On-call Scheduling during a Seasonal Epidemic}

\author[cis,limos]{Omar El-Rifai\corref{mycorrespondingauthor}}
\cortext[mycorrespondingauthor]{Corresponding author}
\ead{el-rifai@emse.fr}

\author[cis,limos]{Thierry Garaix}
\ead{garaix@emse.fr}

\author[cis,limos]{Xiaolan Xie}
\ead{xie@emse.fr}

\address[cis]{CIS, \'Ecole des Mines de Saint-\'Etienne, 158 cours Fauriel CS 62362 42023 Saint \'Etienne, France}
\address[limos]{ROGI, LIMOS UMR CNRS 6158, 63173 Aubi\`ere, France}

\begin{abstract}

Overcrowding in Emergency Departments (EDs) is particularly problematic during seasonal epidemic crises. Each year during this period, EDs set off recourse actions to cope with the increase in workload. Uncertainty in the length and amplitude of epidemics make managerial decisions difficult. We propose in this study a staff allocation model to manage the situation using on-calls. An on-call scheduling policy is proposed to best balance between demand coverage and labor cost under legal constraints of working time. The problem is modelled as a two-stage stochastic Integer Linear Program (ILP) and solved using a Sample Average Approximation (SAA) method. Several epidemic scenarios are defined with data from an ED in Lille, France. 

\end{abstract}

\begin{keyword}
Staff Scheduling \sep Emergency Department \sep On-call \sep Stochastic \sep Optimization \sep SAA \sep Influenza
\end{keyword}

\end{frontmatter}


\section{Introduction}

Demand surges are characteristic of Emergency Departments (EDs) during epidemic crises. The ``white plan'' is an emergency response plan defined in France to meet a sudden increase in activity of a hospital. This plan sets down corrective measures to be undertaken in crises situations. For example, health-care professionals on duty are retained, off-duty personnel are called in gradually and crisis management units are set up. However, the ``white plan'' is costly for hospitals and implies the application of several corrective measures simultaneously. 

Because the ``white plan'' is difficult to put in place, a recurrent phenomenon is observed during seasonal epidemic periods in France. EDs are faced with surges in demand that are difficult to handle but that are not important enough to deploy the ``white plan''.  These peaks are alleviated with empiric ad hoc strategies. For example, when faced with an unexpected increase of patients, the medical staff often stays overtime. These strategies often go unnoticed because no protocol formalizes them and the increase in demand is handled more or less. However, the adjustments can be costly for the hospitals and tiring for the personnel. Furthermore, resilience in the ED is limited and some situations can cause irremediable constraints. An increase in patients' waiting times or a decrease in the quality of service are possible consequences in such situations.

Hospitals spend a large part of their budget on human resources.  It is thus imperative to have a strategy to manage the personnel adequately and even more so during epidemic crises. Some hospitals hire surplus personnel as a recourse to cover the rise in demand during peak seasons. Surplus personnel often need training, an adaptation time and are hired for a long period. For these reasons they are not a very flexible alternative. Additionally, EDs often schedule their permanent workforce relying on past experience. When no formal method is used, there is no guarantee that the resources are going to be deployed appropriately. Deploying resources at the wrong moment is problematic on several levels. When several resources are deployed too soon, the management of the epidemic peak will be difficult in later periods as no more surplus resources will be available. Similarly, if resources are called for too late, the early periods of the epidemic crises become difficult.

In this paper we study scheduling flexibility of the current work force as a way to better meet the seasonal epidemic demand. We analyze the impacts on the costs incurred by the hospital and their ability to better cover the demand.  Staff scheduling is done robustly to account for uncertainties in the evolution of the epidemic. More specifically, we try an alternative to incrementing the workforce by proposing and evaluating an on-call staff allocation policy.  The context of the epidemic season is first analyzed then a staff management strategy to efficiently cope with epidemics is proposed. Efficiency is evaluated with regards to patients' service level and labor cost. A two-stage stochastic programing model is developed to set the problem formally. In the first stage, only partial and incomplete data is available and decisions are made according to estimations of the demand. The second stage involves decisions that are made on a day-to-day basis.

\section{Literature Review}

 Epidemic crises are often studied at the regional level where whole populations and many hospitals interact. Epidemic control and containment include issues ranging from awareness campaigns to the optimal distribution of medicines and vaccines \cite{Flahault2006,Arora2010,Parvin2012,Yarmand2014}. For instance, in a study carried out by Arora et al. \cite{Arora2010} a cost benefit based approach is used for developing a mutual-aid resource allocation strategy between different regions. Several hospitals share resources found in a central ``warehouse'', reminiscence of a warehouse-retailer supply chain. Similarly, in a study by  Yarmand et al.\cite{Yarmand2014}, a central stockpile of vaccines is distributed among different regions using a two-stage stochastic programming approach. However, regional policies can not easily be used in individual hospitals. Particularly, it is difficult to adapt these strategies to a single ED department where the issue is not the distribution of resources across space but across time. 

When considering ED departments independently, the problem of seasonal epidemics is different. Indeed, the main interest of EDs during seasonal epidemics is to effectively handle the increasing ED workload \cite{Bienstock2012,Rico2009}. Few studies have addressed the problem of epidemics from a hospital perspective. It is important to note that during epidemic seasons, ED overcrowding is not strictly due to an increase in the volume of patients. Indeed other factors as the visit duration and acuity of the patients significantly increase ED workload \cite{Sinclair2007, Stang2010}.

Overcrowding affects many people, from the health-care staff to the patients and the people accompanying them.  Resource management strategies to deal with overcrowding involve both human and material resources. The management of these resources relates to decisions on the number, the location and the time in which they should be deployed. As with all service industries, it is difficult for health care industries to make decisions during epidemics because of the uncertainty associated with the workload \cite{Sasser1976}.

In health-care industries, the continuous working environment and the different staff skills make the problem of resources allocation difficult. Consequently, there is an extensive number of articles written on the subject of resources scheduling and rostering \cite{Ernst2004, Burke2004}.From as early as the 1970s, papers \cite{Warner1976, Miller1976} dealt with the problem of nurse scheduling using mathematical programming. In the following years, more complex models accounted for constraints involving different shift types, skill mix, and fairness concerns \cite{Jaumard1998, Beaulieu2000}. As the models complexity increased, the solution methodologies adapted. As such, articles using meta-heuristics and  simulation techniques emerged \cite{Gutjahr2007,Bester2007}. For example, in \cite{Gutjahr2007}, the authors propose an ant colony optimization approach for nurse scheduling so as to include constraints such as nurses' and hospitals' preferences and nurses qualifications. Finally, studies that have addressed the issue of on-call duties in EDs refer for the most part to specialists' on-call duties \cite{Rudkin2004, Menchine2008}.

This study analyzes the situation of seasonal epidemics inside EDs. In particular, it addresses different managerial and operational issues involved during seasonal epidemics. The demand is estimated using a stochastic epidemiological model. We assume that an indicator for the start of the epidemic exists. As such, a threshold of workload inside the ED can be defined beyond which, the system is considered to be overcrowded. After the problem context and assumptions are well defined, a robust ED staff allocation strategy is proposed. The model is a two-stage stochastic linear program solved using a Sample Average Approximation (SSA) approach. The model serves as a basis for the analysis of costs and impacts of different solution strategies. In particular we evaluate the impact of on-call duties on the coverage of epidemic demand. Finally we examine a real case study and compare our solutions with a staff scheduling policy in an ED in Lille France.

\section{Workload Patterns}\label{sec:workload}

\begin{figure}[htp!]
\centering
\includegraphics[width=0.45\textwidth]{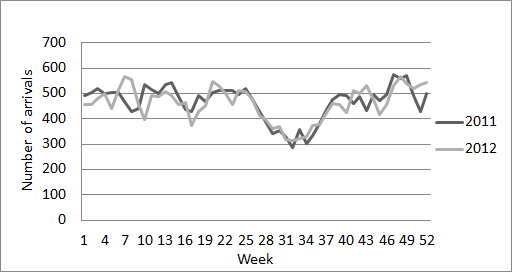}
\caption{Number of patients arriving per week for the years 2011 and 2012}
\label{fig:arrivals}
\end{figure}

The workload inside an ED is fluctuating throughout the year. Usually, during epidemic seasons, hospital personnel experience an increase in the workload. It is generally admitted that this phenomenon occurs annually with varying amplitude. However, if we look at the weekly number of arrivals in Figure \ref{fig:arrivals}, we see that it is not obviously the case.  Indeed, the number of arrivals seems more or less constant throughout the year except during the summer for about two months when there is a decrease in the number of arrivals.

\begin{figure}[htp!]
\centering
\includegraphics[width=0.45\textwidth]{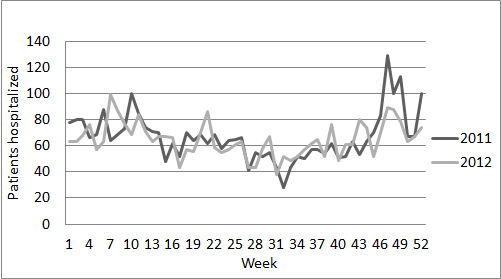}
\caption{Number of patients per week hospitalized after the ED}
\label{fig:hospitalized}
\end{figure}

\begin{figure}[htp!]
\centering
\includegraphics[width=0.45\textwidth]{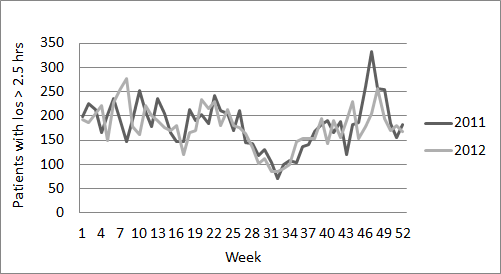}
\caption{Number of patients per week with sojourn time greater than 2.5 hours}
\label{fig:los}
\end{figure}

This phenomenon led us to look into other factors that could influence the workload such as the sojourn time of patients and the destination of the patients after going through the ED. In Figure \ref{fig:hospitalized}, we clearly see that the number of patients hospitalized after their ED visit increases during the last weeks of the year. This is also the case for the sojourn time which increases sharply during the same period as we see in Figure \ref{fig:los}. Consequently, these data corroborate the need for additional capacity during epidemic season.

\begin{figure}[htp!]
\centering
\includegraphics[width=0.45\textwidth]{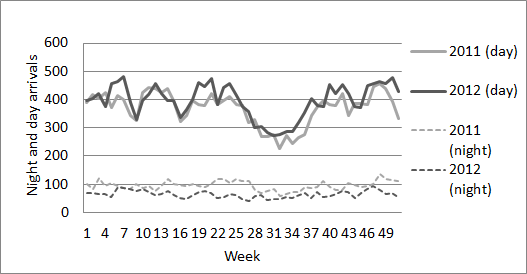}
\caption{Difference of arrivals between night and day}
\label{fig:night_day}
\end{figure}

Once the epidemic horizon is defined, we need to identify trends within it. According to the French law \footnote{http://vosdroits.service-public.fr/particuliers/F573.xhtml}, the work done during the night has different rules than the work done during the day. For work to be considered as night work, it has to be performed during a period including the time from 9 PM to 6 AM. If we look at the number of arrivals to the ED during this night period Figure \ref{fig:night_day}, we notice two things. The night workload is less fluctuating and lower than the day workload.  Consequently, when defining possible epidemic scenarios, the night period needs to be distinguished from the day period.

The number and type of patients in an ED does not directly map to a number of human resources. However, an increase in the workload calls for a parallel increase in resources if a good service quality is to be ensured. It is thus possible to estimate human resources demand given a good estimation of the epidemic development. In Section \ref{sec:allocation} a staff allocation model is presented to adapt the capacity according to the seasonal trends observed. Some assumptions are made in the allocation model and justified accordingly.

\section{Staff Allocation Model}\label{sec:allocation}

 The allocation model we present in this section is proactive in the sense that decision makers take a decision prior to the realization of the epidemic. In reality, this can be achieved by efficient early epidemic declaration mechanisms. The first  point of interest for an ED is to reduce overcrowding and retain a reasonable service level. Patients waiting excessively in EDs before being attended can experience severe health complications. Similarly, physicians can experience burnouts as a result of excessively stressful work patterns. From a hospital point of view, it is important to minimize the cost associated with recourse actions. The problem is not trivial because of financial and operational constraints and the stochastic nature of the workload. 

 As mentioned earlier, when there is excess workload, a natural tendency of care-practitioners is to modify their attitude and adapt to stressful situations at the expense of their own well being. The problem of optimal human resource allocation is then an actual concern in EDs during seasonal epidemics. However, extending the capacity of EDs is not easily implementable. A human resource management strategy is thus crucial for efficient service in EDs.

In order to develop a mathematical model suitable for the problem of staff allocation during epidemic crises, some assumptions are made. First we notice that during epidemics, the staff can also be exposed to infections. But as we are interested in seasonal epidemic that display mild infection patterns, we assume that the effects are negligable. In practice, even if a certain particular health-care practitioner is indisposed, managers find substitutes to replace him, albeit at a greater cost.  Modeling the availability of  staff makes sense in the case of large pandemics where the majority of the population is at risk.

Then, only one class of patients is considered. We assume that these patients are homogeneous in visit length, discharge and resources needed. A look at the data from the UHC of lille shows that during epidemic seasons, the arrivals are primarily from epidemic patients. This peak in arrivals generates additional workload as these epidemic patients spend more time in the ED than the rest of the year. Moreover, patients' destination following their visit is not important insofar as the resources deployed will not be responsible for the patients after their visit.

Another assumption is that based on the available information at the start of the epidemic, decision makers can characterize the resources needed across the planning horizon. Of course, these estimations are subject to errors and represented as random variables but the general trend can be forecast.  In reality , at the time of the declaration of the epidemic, the strain of the virus is known and epidemiological models can be used to simulate the spread of the disease. Specifically, in this study we do not deal with special cases of epidemics where new viruses cause a pandemic. Rather, we examine cases of periodic overcrowding resulting from well studied pathologies such as known Influenza strains. These simulation models estimate the number of infected persons in a population per time period. It is then reasonable that ED decision-makers anticipate with a margin of error the number of resources required across the planning horizon.

We also make the assumptions that resources can be scheduled independently. This implies that given a solution to the problem proposed, more operational decisions such a shift assignment with resource interactions and different patient flows are easier to take. The assignment of work periods to employees defines the day (or night) in which the employee is on duty but does not specify the type of shift he/she is on. We assume that given the 12 hours period in which an employee is working, it is always possible to come up with a shift assignment that satisfies individual constraints. This is possible if individual constraints are already ensured at the tactical level as we did in this study. Furthermore, since the workload remaining from a day to the other is very low because patients can not stay more than a day in EDs, the workload in the early morning is almost null and there is no accumulation of work.

\subsection{On-Call Duties}

Employees in France have one of three statuses: working, resting or on-call. This peculiar third status is defined as a period during which employees must be ready to intervene at work without being at the permanent and immediate disposal of the employer. When an employee is called in, the intervention periods including the transportation time are considered effective work hours. Otherwise, employees stay at home or in the vicinities and are paid a percentage of their hourly wage. 

On-call duties are not widely applied in ED yet. This reluctance to use on-calls comes from both practitioners and managers. For practitioners, on-call duties are associated with extra work hours as labor laws are sometimes circumvented. Managers on the other hand are put off by the strictness of the labor laws. Specialized physicians are an exception as they often have on-call duties in the ED while working primarily in another department. However, according to a public service decree ``The head of the hospital, after an agreement with the technical committee of the hospital, determines the list of activities, services and jobs that are concerned [with on-call duties]''. Furthermore, the law specifies the modalities of work for public service employees.  When an employee intervenes while on-call, on-call duties turn into effective work hours. Consequently, the normal work regulations apply.

The on-call payment scheme is beneficial for both employees and employers. Employers have staff on demand paying them a fraction of their hourly wage and employees are paid while staying at home. Additionally, if they called in to work, they are paid extra hours. Although flexible, this organizational scheme is not currently used in emergency departments in France.  This study analyzes the advantages of allowing for employees to be on-call during epidemic crises. For doing so, a comparison is made with a regular ED schedule.

\subsection{Mathematical Formulation} \label{sec:math_form}

When an epidemic starts, ED managers know that demand will increase but have little or no information on the amplitude and duration of the epidemic. For that reason, it is logical to have some flexibility in the staff deployment plan. Decisions are made in two stages. The first stage decision is made at the start of the epidemic and consists of contracting resources to be on-call or on a regular duty in period $t$. Second stage decisions are made successively for each period and determine the number of resources to call for work among the ones that are on-call.

 When on-call, physicians have a fixed hourly compensation $w$ regardless of their actual wage. If these physicians are called in to work they are paid at overtime rates $d$. The distribution of the resources $n$ among the $T$ half day periods needs to respect the working time constraints imposed by the law. The rest of the employees is ensured by variables $U_n$ that keep track of the worked periods in a week. Also parameters $O_n$  impose an upper bound on the on-call duties possible and parameters $H_n$ impose an upper bound on the night shifts allowed per resource $n$. The resources that can be allocated to on-call duties are those that are not on a regular shift during that day. 

Consider the total number of patients arriving to the ED in period $t = 1,..,T$. Once this random process is realized, it is possible to estimate the number of physicians required to meet this workload. In our model, for each realization $\xi$, we denote $b_{nt}(\xi)$ the number of physicians to meet this demand. At the beginning of the planning horizon  we take the decisions $x_{nt}$ and $y_{nt}$ to allocate physicians respectively on duty or on-call for periods $t \in 1..T$. The decisions are robust insofar as they are optimized with respects to several possible epidemic realizations. We denote the matrices containing the variables $x_{nt}$ and $y_{nt}$ by $x$ and $y$ respectively.

If a resource is scheduled to be on-call during period $t$, a decision $y'_{n,t}$ whether to call resource $n$ is made for each period. This decision depends on a compromise between paying an extra fee or incurring a penalty $\alpha$ for not meeting the demand. Tables \ref{tab:variables} and \ref{tab:params} summarize the different parameters and variables of the model.

\begin{table}
\centering
\begin{tabular}{c | p{10cm} }
Parameter & Description \\
\hline
$T$ & Length of the planning period  indexed by $t$\\
$N$ & Total number of physicians indexed by $n$\\
$M$ &  Minimal number of physicians on duty by period\\
$b_t$ & Number of physicians needed in period $t$ \\ 
$c$ & Cost to put an employee on duty by period\\
$w$ & Cost to put an employee on-call by period \\
$d$ & Cost to call in an on-call personnel  \\
$W_n$ & Minimal number of duty periods of a physician $n$ \\
$O_n$ & Maximal number of on-call periods of $n$\\
$N_t$ & Minimal number of on-duty physicians in period $t$ \\
$H_n$ & Maximal number of night shifts allowed for resource $n$ \\
$\alpha$ & Shortage cost coefficient \\
\end{tabular}
\caption{Parameters of the model.}
\label{tab:params}
\end{table}

\begin{table}
\centering
\begin{tabular}{c | p{10cm} }
Variable & Description \\
\hline
$x_{nt}$ & = 1 if physician $n$ is on duty in period $t$  \\
$y_{nt}$ & = 1 if physician $n$ is on-call in period $t$  \\
$y'_{nt}(\xi)$ & = 1 if physician $n$ is called in period $t$ \\
$U_{nt}$ & = 1 if physician $n$ works at least one period during the three periods following t $t$ \\ 
$l_{t}(\xi)$ & Physician shortage in period $t$ \\
\end{tabular}
\caption{Variables of the model}
\label{tab:variables}
\end{table}

\begin{center}
 {\bf Objective:}
\end{center}

\begin{equation}
min \hspace{0.5cm} \sum_t \sum_n (c * x_{nt} + w * y_{nt}) + \mathbb{E}[Q(x,y,\xi)]
\label{obj}
\end{equation}

\begin{center}
{\bf Subject to:}
\end{center}

\begin{equation}
W_n \leq \sum_t x_{nt}  \hspace{0.5cm} \forall n
\label{c:xupper}
\end{equation}

\begin{equation}
\sum_t y_{nt} \leq O_{n} \hspace{0.5cm} \forall n 
\label{c:yupper}
\end{equation}

\begin{equation}
\sum_{t \in night} (x_{nt} + y_{nt}) \leq H_{n} \hspace{0.5cm} \forall n
\label{c:night_limit}
\end{equation}

\begin{equation}
x_{nt+1} + y_{nt-1} + x_{n,t} + y_{n,t}  \leq 1 \hspace{0.5cm} \forall n \hspace{0.25cm} \forall t
\label{c:consecutive}
\end{equation}

\begin{equation}
\sum_{t'=t}^{min(t+2,T)} (x_{nt'} + y_{nt'})  \leq 3 U_{nt} \hspace{0.5cm} \forall n \hspace{0.25cm}  \forall t
\label{c:consecutive_week1}
\end{equation}

\begin{equation}
\sum_{t'=t}^{min(t+11,T)} U_{nt'} \leq 11 \hspace{0.5cm} \forall n \hspace{0.25cm} \forall t
\label{c:consecutive_week2}
\end{equation}

\begin{equation}
\sum_{n} x_{nt}  \geq M \hspace{0.5cm} \forall t
\label{c:lowerbound_assigned}
\end{equation}

\begin{equation}
U_{nt}, x_{nt}, y_{nt} \in \{0,1\} \hspace{0.5cm} \forall n, \hspace{0.25cm} \forall t 
\label{c:int}
\end{equation}

$Q(x,y,\xi)$ denotes the solution to the second stage problem where $Y'_t = \sum_n y'_{nt}$:

\begin{equation}
Q(x,y,\xi) =  min  \sum_t  d * Y'_{t}(\xi) + \alpha l_t(\xi)
\label{obj:2}
\end{equation}

\begin{equation}
\sum_n y_{nt} \geq Y'_{t}(\xi)   \hspace{0.5cm} \forall t \hspace{0.25cm} \forall n \hspace{0.25cm}
\label{c:xassign}
\end{equation}

\begin{equation}
 b_{t}(\xi) - Y'_{t}(\xi) - \sum_n x_{nt} \leq l_t(\xi)\hspace{0.5cm} \forall t \hspace{0.25cm}
\label{c:demand}
\end{equation}

\begin{equation}
l_t(\xi) \geq 0 \hspace{0.5cm} \forall t
\label{c:llower}
\end{equation}



Planners are interested in obtaining a robust allocation scheme in the beginning of the planning horizon. The objective function \eqref{obj} tries to find such a scheme by minimizing the total cost over the planning horizon while trying to meet as much demand as possible. If EDs are overcrowded, i.e if the demand is above the service capacity, there are several consequences with regards to patients. Besides possible health complications, patients can both be delayed and leave without being seen. However, as we are dealing with decisions that improve patients throughput, it is much more likely in this context that patients remain in the ED even with increased waiting times. Patient's outcome is not explicitely stated, but an indicator of the amount of expected shortage is meaningful for decision makers who can relate to situations when resources were missing.  We are interested in evaluating the impact of on-calls on schedules costs. For this reason, we assign a cost to regular duties as well as to on-call duties even if employees are no paid by duty but have a fixed salary.

 The first set of constraints \eqref{c:xupper} ensure resources have to work on regular duties for at least $W_n$ periods. This prevents the allocation model from assigning only on-call duties. Constraints \eqref{c:yupper} set an upper bound on the number of on-call duties that are assigned to the resources. Similarly, this prevents the allocation model from assigning more on-call duties than what is legally allowable. Next, constraints \eqref{c:night_limit} set an upper bound on the number of night shifts that employees can have during the epidemic horizon. Since these limit depends on the work history, contract or preferences of each individual employee, they are indexed by $n$. Constraints \eqref{c:consecutive} state that an employee can not be assigned to a regular duty and an on-call duty on the same period. Furthermore, an employee can not be assigned to duties on two consecutive periods. Admittedly the constraints on the on-call duties are over conservative as on-call duties are counted as effective work hours only if an employee is called. The actual constraints as imposed by law involve the variables $y'_{nt}$ rather than $y_{tn}$. However, since the variable $y'_{nt}$ refer to decisions that are scenario-dependent, having a constraint on these over optimizes the solution as if the future realizations were known while taking the decisions. Constraints \eqref{c:consecutive_week1} and \eqref{c:consecutive_week2} state that employees must rest for at least 3 consecutive periods in a week. This rest corresponds to the 36 hours of weekly rest imposed by the labor law. We first go through all three consecutive periods and verify if one of those periods is a work period. If we find three consecutive rest periods (i.e $x_{nt'}$ and $y_{nt'}$ are zero) then the variable $U_{nt}$ can be assigned 0. Otherwise the value of $U_{nt}$ is 1. Constraints \eqref{c:consecutive_week2} then checks that there is at least one $t$ in the following week (14 periods) where $U_{nt}$ is not 1.  Finally, constraints \eqref{c:int} ensure that the variables $x_{nt}$, $y_{nt}$ and $U_{nt}$ are binary.

The second stage objective \eqref{obj:2} is to find the optimal number of resources to call for each epidemic scenario given there is a penalty on the resource shortage.  Constraints \eqref{c:xassign} are logical constraints that make sure that only employees that are on-call can be called. Constraints \eqref{c:demand} are demand satisfaction constraints with the variable $l_t(\xi)$ keeping track of the shortage on period $t$. The variable $l_{t}(\xi)$ is calculated from a random number of infected patients. In the model, an equivalent formulation is to have the demand in number of patients denoted as $b_t(\xi)$ and a constraint specifying:

\begin{equation}
l_{t}(\xi) = b_t(\xi) * \beta
\end{equation} 

With $\beta$ being the ratio of resources to patients that we can calculate based on information on the epidemic and physician's experience. The ratio $\beta$ in our tests is a random variable itself because it depends on individual patients and the service rate. We chose to hide this equation and generate data independently leaving the possibility of another mean of demand generation. We know that queuing effect are negligible from a day to another since the workload at night is almost always completely absorbed. Finally, constraints \eqref{c:llower} are positivity constraints for variables $l_t(\xi)$. We can show that the second stage objective reduces to the following two cases:

\begin{align}
\begin{cases}
Q(x,y,\xi) = \alpha l_t(\xi) & \mbox{if } d > \alpha \\
Q(x,y,\xi) = \alpha l_t(\xi) + d Y'_t(\xi)  & \mbox{if } d <= \alpha
\end{cases}
\end{align}

 In reality, in order to mirror the day-to-day behavior where decisions are taken without an accurate vision of the future, non-anticipativity constraints need to be satisfied. Optimal second stage decisions $Y'_{t}(\xi)$  are null if $d > \alpha$ and $Y'_{t} = \min( b_t (\xi), \sum_n y_{nt})$ otherwise. Consequently, these decisions are independent from future realizations.

\section{Numerical Experiments}

In the following set of tests we show the advantages and shortcomings of using alternative scheduling strategies and analyze whether they help us handle uncertainties during an epidemic season. All tests were run on an Intel(R) Xeon(R) 4 cores CPU E5520 @ 2.27GHz with 8 MB of cache memory and 8 GB of RAM. We choose a Sample Average Approximation method (SAA) approach to solve the model \cite{Kleywegt2002}. The Sample Average Approximation (SAA) method is a solution methodology for stochastic discrete optimization problems based on random sampling of data from the input space. Random samples are generated and the expected value function is approximated by the corresponding sample average function. With the increase of sample size, the solution converges to the real optimum. The sample size is chosen to guarantee an acceptable margin of error. We used samples estimated based on epidemic simulations. After we parametrize the epidemiological model based on the epidemic data of interest, we calculate a number of infected persons in a population and deduce the expected demand from those. In Table \ref{tab:stability} the stability of the model is tested for different number of samples realization. The column Obj refers to the average  value of the objective function obtained from three set of realizations in each case.  As the objective value seems to stabilize around 100 samples, this number is used throughout for the remaining tests.

Unless otherwise stated, the default values of the parameters used for the tests are shown in Table \ref{tab:parameters}. The choice of the cost parameters are motivated by labor laws stating that the cost of an on-call duty is $1/4$ the cost of a regular duty if the employee is not called in for work. Otherwise the employee is paid extra hours. Furthermore, the value of the shortage parameter $\alpha$ was chosen so as to be higher than the cost of any work duty. This is consistent with equation (14) which states that otherwise the solution to the problem is trivial. We picked a 30 days planning horizon which corresponds more or less to the length of an epidemic based on physicians' experience. The number of on-call duties and night shifts are motivated by labor laws and illustrate a realizable set up given the  length of the planning horizon. Finally, the value of $M$ states that there should be at least one physician present at all times in the ED. This parameter can be adjusted but serves to illustrate one practical example.

\begin{table}[h!]
\centering
\begin{tabular}{ c c }
Parameter & Value \\
 c & 4 \\ 
 w & 1 \\ 
 d & 4 \\
$\alpha$ & 7 \\ 
T & 60 \\
I & 13 \\
W & 10 \\
$O_n$ & 10 \\
H & 10 \\
M & 1\\
\end{tabular}
\caption{Model Parameters}
\label{tab:parameters}
\end{table}

\begin{table}[htbp]
\centering
\begin{tabular}{c c}
$\Xi$ & Obj \\ \hline
50 & 1054.5 \\
100 & 1070.9 \\
200 & 1075.0 \\
500 & 1064.3 \\
1000 & 1068.4 \\
\end{tabular}
\caption{Numerical stability through different number of samples}
\label{tab:stability}
\end{table}

\subsection{Design of Experiments}

In order to examine the impact of a flexible planning strategy in facing epidemic cases, the cost of different planning alternatives are studied. In Section \ref{sec:scenarios} we examine three types of epidemics: mild, moderate and severe. These scenarios serve to estimate the arrivals in the ED and consequently the number of resources needed. Furthermore, the shortage cost is a key parameter in our model as it corresponds to a penalty cost for being short in capacity. As the value of this penalty cost can not easily be estimated we investigate in our tests schedules with the value of $\alpha$ in a range of +20\% to +500\% of the cost of an on-call duty.

 In Section \ref{sec:cyclic} the impact of on-call duties on ED schedules is studied. To do that, we impose a weekly cyclic schedule constraint on regular duties and allow on-call duties to be distributed optimally. The cyclic schedule  where the number of resources per period is fixed mimics the current scheduling practice. The results are compared to a schedule with no on-call duties. Next, in Section \ref{sec:non_cyclic}  we relax the cyclic constraints and analyze whether on-calls are still beneficial in the presence of non-cyclic duties. Having non-cyclic schedules makes sense in the case accurate demand forecast is available.

The French laws for on-call duties are restrictive as they do not lead to an increase in the hospital's work capacity. Indeed, whether employees are on-call or not, the total work hours remains unchanged. In that sense, the only advantage of on-call duties is in the managerial flexibility they provide compared to regular duties. In practice, some hospitals use on-call duties as a surge mechanism in exceptional circumstances. Instead of counting work performed by employees who were on-call as effective work hours, they consider it to be paid ``extra work''. Of course, doing so means they bypass the law and effectively increment their capacity. We examine in Section \ref{sec:no_constraints} how the optimal distribution of duties differs in the case laws are circumvented.

\subsection{Demand Generation}\label{sec:scenarios}

 Influenza-like-illnesses (ILI) or acute respirator infection (ARI) are umbrella terms that refer to medical disorders whose symptoms resemble those of Influenza cases. These concepts are important inasmuch as they allow us to group certain pathologies and study their collective incidence on EDs. During epidemic seasons, ILI are the main sources of ED activity. Patients with disease like influenza, bronchiolitis and gastroenteritis add up to the regular volume of patients and significantly increase the workload. As we have said previously, the increase in workload is not primarily due to an increase in patients' volume but rather in an increase in the time spent in the ED and acuity of the patients visiting.

From year to year, epidemics differ in nature. This uncertainty leads to managerial complexities. EDs struggle to  characterize the seasonal workload peak so as to organize their service and serve the demand better. From a modeling perspective, this uncertainty has to be taken into account if the model is to solve the difficulties encountered in the ED. There is uncertainty as to the type of epidemic and as to the intensity and fluctuation within one type of epidemic. A stochastic SEIR model is used to simulate realizations of the epidemic with the three different contact rates. 

\begin{figure*}
\centering
\begin{subfigure}[b]{0.3\textwidth}
\includegraphics[width = \textwidth]{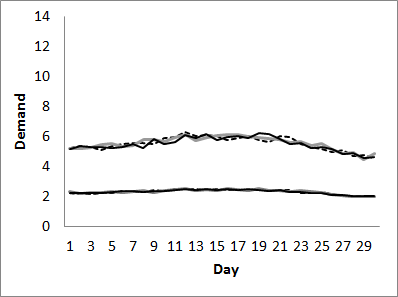}
\caption{mild}
\label{fig:mild}
\end{subfigure}
\begin{subfigure}[b]{0.3\textwidth}
\includegraphics[width = \textwidth]{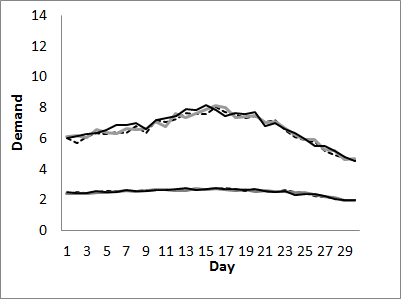}
\caption{moderate}
\end{subfigure}
\begin{subfigure}[b]{0.3\textwidth}
\includegraphics[width = \textwidth]{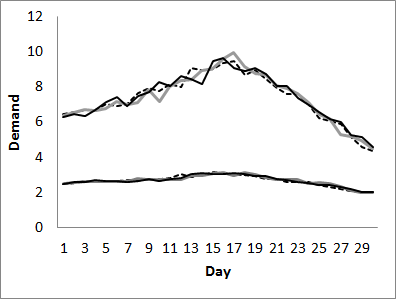}
\caption{severe}
\end{subfigure}
\caption{Day and night demand for three types of epidemic scenarios}
\label{fig:epidemics}
\end{figure*}

Running experiments with three different contact rates illustrate the solutions in different epidemic scenarios (mild = 0.4, moderate = 0.6 and severe = 0.8). In practice, the contact rate of an epidemic can be estimated by specialists if the type of the epidemic is known. The epidemic model for simulating the spread of disease is represented as a stochastic Markov chain with four compartments: Susceptible, Exposed, Infectious and Recovered which record the number of individuals in each category. The initial state has few infected individuals. At each time step (fraction of a minute), there is a probability that an individual goes to another state in the system. We chose to use these models to generate demand because they provide good approximations of the increase and reduction of infected population across time. As we are primarily interested in the workload associated with infected patients, we believe that estimating demand based on an epidemiological model provides good basis for planning.  In this study, we are interested in the evolution of the number of individuals in the infectious category \cite{Chowell2008}. Parameters of the SEIR model are fitted to the population of the area around the ED. Figure \ref{fig:epidemics} shows the average number of required resources calculated for each type of epidemic for both day and night. The averages are calculated from 3 instances of 100 sample realizations each. Demand at night is lower than during the day and does not follow the epidemic as explained in Section \ref {sec:workload}.

 Based on the number of infectious individuals in a population, ED arrivals and resources needed are calibrated for a particular hospital  in Lille France. Consequently, the number of resources used for our tests allow us to cover in average all the demand of Figure \ref{fig:mild} without any shortage.The three epidemic scenarios and the varying values of $\alpha$ are at the base of all the following experiments forming 15 test instances. We start in the next section by testing the benefits of on-calls compared to current scheduling practice.

\subsection{Cyclic Schedules}\label{sec:cyclic}
So as to evaluate the impact of on-calls on the scheduling costs, we compare schedules with and without on-calls. In both cases, as in current practice, the schedules of regular duties are cyclic. This means that they are optimized for one week and repeated for the rest of the planning horizon.  Table \ref{tab:cyclic_with} and \ref{tab:cyclic_without} show the results obtained for the moderate epidemic scenario.

 The first column shows the percentage added to the cost $\alpha$ compared to the cost of an on-call duty. For example, for $\alpha = +20\%$ the penalty cost for the shortage is set to $w+d+((w+d)*20/100)= 6$ . For each value of $\alpha$, three instances are run each with 100 sample epidemic realizations.  All tests that we ran finished in less than an hour for a relative optimality gap that was set to 0.01\%. The next column shows the time in seconds to obtain the solution.  After that the column ``Fix'' specifies the number of regular duties assigned in the optimal solution. Then ``On-call'' specifies the number of on-call duties assigned. Finally, the columns ``Exp. Calls''' and ``Exp. Short'' show respectively the expected number of calls made to personnel that were on-call and the expected number of shortage across the horizon.

\begin{table}[htbp]
\centering
\begin{tabular}{|c|c|c|c|c|c|c|c|}
\hline
{\bf Scenario} & {\bf $\alpha$} & {\bf Obj} & {\bf Time (s)} & {\bf Fix} & {\bf On Calls} & {\bf Exp. Calls} & {\bf Exp Short} \\ \hline
\multirow{5}{*}{mild} & 1- +20 \% & 1055.8 & 57.8 & 191 & 43.3 & 28 $\pm$ 3.7 & 22.3 $\pm$ 4.6 \\ 
& 2- +60 \% & 1075.8 & 122.2 & 191 & 89.7 & 45.3 $\pm$ 6.1 & 4.3 $\pm$ 2.1 \\ 
& 3- +100 \% & 1081 & 204 & 191 & 104.7 & 49 $\pm$ 6.6 & 1 $\pm$ 1.3 \\ 
& 4- +200 \% & 1085.9 & 241.5 & 191 & 110.3 & 49.3 $\pm$ 6.9 & 0.3 $\pm$ 0.9 \\ 
& 5- +500 \% & 1090.1 & 280.1 & 191 & 116 & 50 $\pm$ 6.9 & 0 $\pm$ 0.5 \\ \hline \hline
\multirow{5}{*}{moderate} & 1- +20 \% & 1254.3 & 137.6 & 209.7 & 77.7 & 51.3 $\pm$ 5.3 & 21 $\pm$ 4.8 \\ 
& 2- +60 \% & 1281.4 & 695.2 & 211 & 102 & 61 $\pm$ 6.6 & 10.7 $\pm$ 3.5 \\ 
& 3- +100 \% & 1296.1 & 1095.4 & 212.7 & 107.7 & 61.7 $\pm$ 6.8 & 8.7 $\pm$ 3.2 \\ 
& 4- +200 \% & 1330.2 & 1540.6 & 212.7 & 114.7 & 63.3 $\pm$ 7 & 6.7 $\pm$ 3 \\ 
& 5- +500 \% & 1433.8 & 2072.4 & 209.3 & 119.3 & 66.3 $\pm$ 7.2 & 6.3 $\pm$ 3 \\ \hline \hline
\multirow{5}{*}{severe} & 1- +20 \% & 1389.4 & 469.2 & 218 & 81 & 57.3 $\pm$ 5 & 33.3 $\pm$ 6.4 \\ 
& 2- +60 \% & 1438.3 & 1105.8 & 222 & 98.7 & 64.7 $\pm$ 5.7 & 23.3 $\pm$ 5.6 \\ 
& 3- +100 \% & 1481.9 & 1375.5 & 222 & 104.3 & 66.3 $\pm$ 6 & 21.7 $\pm$ 5.6 \\ 
& 4- +200 \% & 1585.6 & 1727.2 & 224.7 & 104.3 & 65.3 $\pm$ 5.9 & 21 $\pm$ 5.5 \\ 
& 5- +500 \% & 1900.5 & 1673.1 & 226 & 104.7 & 64.3 $\pm$ 5.9 & 20.3 $\pm$ 5.5 \\ \hline
\end{tabular}
\caption{Results for cyclic schedules with on-call duties}
\label{tab:cyclic_with}
\end{table}

\begin{table}[htbp]
\centering
\begin{tabular}{|c|c|c|c|c|c|c|l|}
\hline
{\bf Scenario} & {\bf $\alpha$} & {\bf Obj} & {\bf Time (s)} & {\bf Fix} & {\bf On Calls} & {\bf Exp. Calls} & {\bf Exp Short} \\ \hline
\multirow{5}{*}{mild} & 1- +20 \% & 1062.3 & 40.3 & 210.0 & 0.0 & 0.0 & 36.3 $\pm$ 6.1 \\ 
& 2- +60 \% & 1124.9 & 43.6 & 236.7 & 0.0 & 0.0 & 21.7 $\pm$ 4.5 \\ 
& 3- +100 \% & 1167.1 & 44.1 & 240.0 & 0.0 & 0.0 & 20.3 $\pm$ 4.3 \\ 
& 4- +200 \% & 1232.3 & 45.5 & 291.7 & 0.0 & 0.0 & 3.7 $\pm$ 2 \\ 
& 5- +500 \% & 1269.7 & 42.8 & 300.0 & 0.0 & 0.0 & 2 $\pm$ 1.4 \\ \hline \hline
\multirow{5}{*}{moderate}& 1- +20 \% & 1271.9 & 53.5 & 235.0 & 0.0 & 0.0 & 54.7 $\pm$ 7.5 \\ 
& 2- +60 \% & 1358.7 & 55.7 & 281.0 & 0.0 & 0.0 & 28.7 $\pm$ 5.7 \\ 
& 3- +100 \% & 1399.8 & 59.3 & 308.0 & 0.0 & 0.0 & 16.3 $\pm$ 4.6 \\
& 4- +200 \% & 1460.8 & 58.4 & 324.0 & 0.0 & 0.0 & 10.7 $\pm$ 3.7 \\
& 5- +500 \% & 1566.5 & 311.5 & 338.3 & 0.0 & 0.0 & 6.3 $\pm$ 3 \\ \hline \hline
\multirow{3}{*}{severe} & 1- +20 \% & 1414.2 & 51.6 & 249.0 & 0.0 & 0.0 & 69 $\pm$ 8.8 \\ 
& 2- +60 \% & 1509.4 & 59.6 & 312.0 & 0.0 & 0.0 & 32.3 $\pm$ 6.7 \\ 
& 3- +100 \% & 1567.5 & 61.1 & 327.0 & 0.0 & 0.0 & 25.7 $\pm$ 5.9 \\
& 4- +200 \% & 1676.4 & 265.5 & 336.7 & 0.0 & 0.0 & 21.7 $\pm$ 5.4 \\
& 5- +500 \% & 1996.1 & 501.2 & 339.0 & 0.0 & 0.0 & 20.7 $\pm$ 5.4 \\ \hline
\end{tabular}
\caption{Results for cyclic schedules without on-call duties}
\label{tab:cyclic_without}
\end{table}

\begin{figure}[htpb]
\centering
\includegraphics[width=0.49\textwidth]{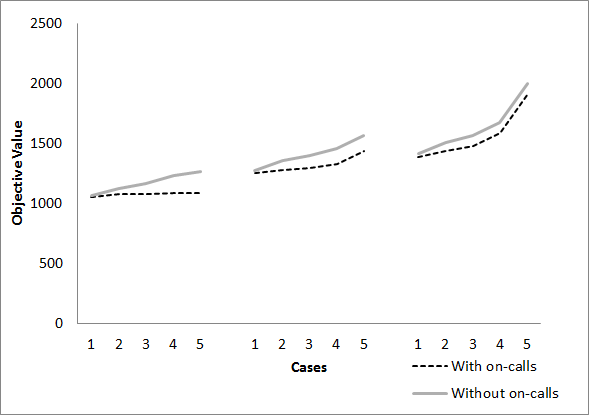}
\caption{Difference in objective value with and without on-calls for different values of $\alpha$ in the three scenarios}
\label{fig:cyclic}
\end{figure}

\begin{figure*}
\centering
\begin{subfigure}[b]{0.45\textwidth}
\includegraphics[width=0.99\textwidth]{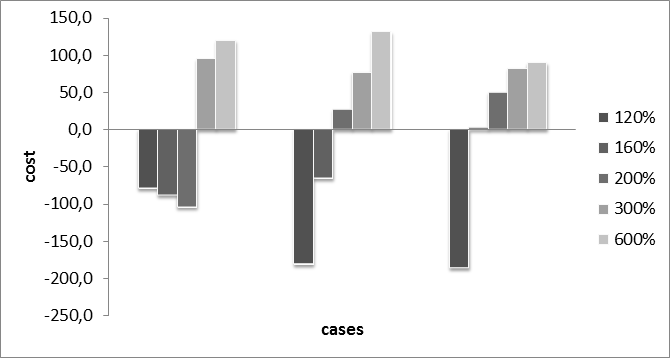}
\caption{Financial cost reduction with on-call vs without on-call}
\label{fig:diff_cost_cyc}
\end{subfigure}
\begin{subfigure}[b]{0.45\textwidth}
\includegraphics[width=0.99\textwidth]{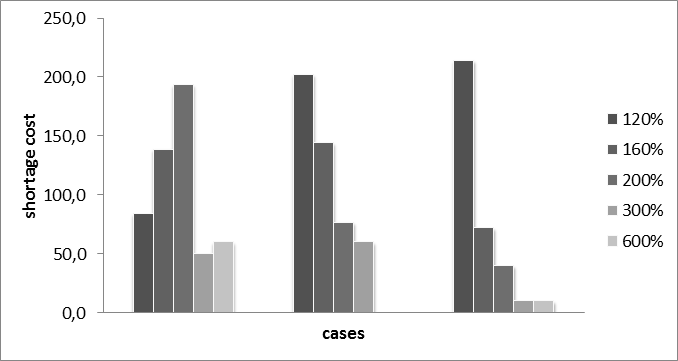}
\caption{Shortage cost reduction with on-call vs without on-call}
\label{fig:diff_short_cyc}
\end{subfigure}
\caption{Costs of cyclic schedules in the mild, moderate and severe scenarios}
\end{figure*}

 Figure \ref{fig:cyclic} illustrates the difference in the objective value obtained for schedules with and without on-call duties for the three scenarios. We see that for all cases, the schedules with on-call duties dominate the schedules without. The values on the x-axis designate the different test cases (different values of $\alpha$.  So as to better understand how the plannings are configured we analyze the difference in costs for the objective functions. Figure \ref{fig:diff_cost_cyc} and Figure \ref{fig:diff_short_cyc} show the difference in respectively the financial and shortage costs with and without on-calls.  A positive difference means that the schedules with on-calls are more expensive.

 When the penalty on the shortage $\alpha$ is low, the schedules without on-calls fail to cover the demand as adequately as the schedules with on-call. For example for $\alpha = +20\%$ There is an expected shortage of 54.7 resource-periods for the moderate epidemic scenario in the schedules without on-calls against 21 resource-periods in the schedules with on-calls. As the value of $\alpha$ increase, the difference in shortage is reduced but the schedules with on-call gain on financial costs. This difference in cost is due to the fact that on-call duties only cost a fraction of regular duties when they are not used. As such, only a fraction is paid for several epidemic realizations while covering the demand in the other cases. Financial cost is calculated by multiplying the number of effective work hours by the unit cost of a work hour.

\begin{figure}
\centering
\includegraphics[width=0.45\textwidth]{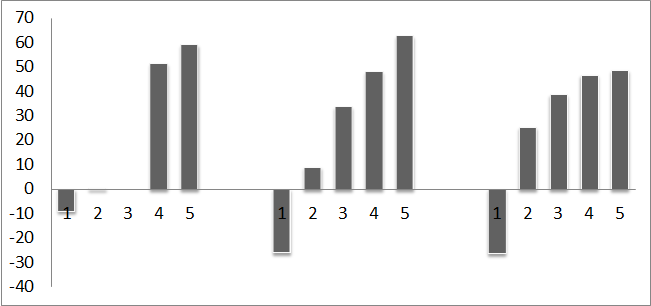}
\caption{Effective work reduction with on-call vs without on-call for cyclic schedules}
\label{fig:diff_effective}
\end{figure}

To further understand the differences in the schedules obtained, we plot in Figure \ref{fig:diff_effective} the difference in actual worked hours between the schedules without and with on-calls. A positive difference means that the the schedules without on-calls have more effective worked hours. The difference is simply calculated as: $x [no\hspace{0.1cm}on-call] - (x [on-call] + \mathbb{E}(Y'[on-call]))$. As the value of $\alpha$ and the number of on-call duties increase, employees work less effective hours. In other words, schedules with on-call duties are more likely to lead to more rest for the employees.

\subsection{Non-cyclic Schedules}\label{sec:non_cyclic}

\begin{figure}[htpb]
\centering
\includegraphics[width=0.49\textwidth]{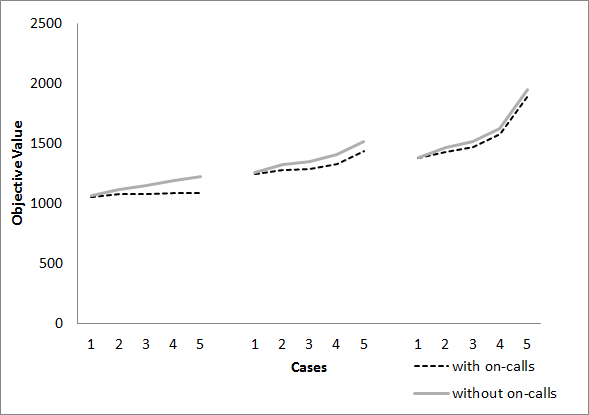}
\caption{
Difference in objective value with and without on-calls for non-cyclic schedules}
\label{fig:obj_non}
\end{figure}

In this section, we relax the cyclic constraints on the schedules. Figure \ref{fig:obj_non} shows the resulting objective values obtained for schedules with and without on-call duties again. The values we observe are very similar to the costs obtained when the cyclic constraints are imposed.

\begin{figure*}
\centering
\begin{subfigure}[b]{0.45\textwidth}
\includegraphics[width=0.99\textwidth]{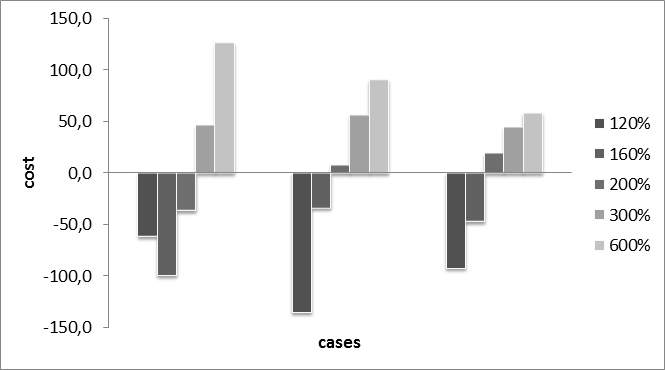}
\caption{Financial cost reduction with on-call vs without on-call}
\label{fig:diff_cost_non}
\end{subfigure}
\begin{subfigure}[b]{0.45\textwidth}
\includegraphics[width=0.99\textwidth]{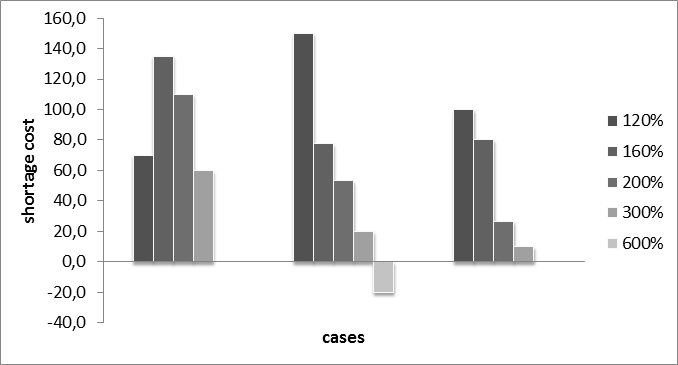}
\caption{Shortage cost reduction with on-call vs without on-call}
\label{fig:diff_short_non}
\end{subfigure}
\caption{Costs of non-cyclic schedules in the mild, moderate and severe scenarios}
\end{figure*}

If we examine the details of the costs in the schedule in Figure \ref{fig:diff_cost_non} and \ref{fig:diff_short_non}, we observe a similar behavior than the one in Section \ref{sec:cyclic}. When the values of $\alpha$ is low, the schedules without on-calls do not cover the demand as well as the schedules with on-calls. As $\alpha$ increases, the coverage becomes similar but the schedules with on-calls are cheaper. The results suggest that relaxing the cyclic constraint do not bring a lot of benefit over having optimized cyclic schedules.

\begin{figure}
\centering
\includegraphics[width=0.45\textwidth]{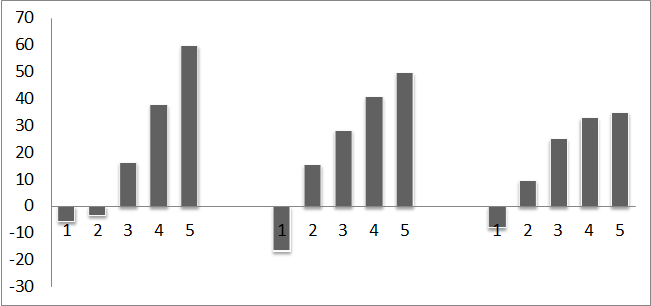}
\caption{Effective work reduction with on-call vs without on-call for non-cyclic schedules}
\label{fig:diff_effective_noncyclic}
\end{figure}

Furthermore, in Figure \ref{fig:diff_effective_noncyclic} a similar pattern than the one observed in Figure \ref{fig:diff_effective} is seen. This similarity in the distribution of work duties suggests that the flexibility allowed by the regular duties are already exhausted in optimizing in a cyclic manner.

\subsection{Relaxed Constraints on Workload Regulations}\label{sec:no_constraints}

In this section, we relax the weekly rest constraints relating to on-call duties and study the effect on the financial cost and shortage costs of the resulting schedules.

\begin{table}[htbp]
\centering
\begin{tabular}{|c|c|c|c|c|c|l|l|}
\hline
{\bf Scenario} & {\bf $\alpha$} & {\bf Obj} & {\bf Time (s)} & {\bf Fix} & {\bf On Calls} & {\bf Exp. Calls} & {\bf Exp Short} \\ \hline
\multirow{5}{*}{moderate} & 1- +20 \% &  1248.7 & 69.3 & 224.0 & 66.3 & 41.3 $\pm$ 4.8 & 19.3 $\pm$ 4.5 \\
& 2- +60 \% & 1266.3 & 127.2 & 222.0 & 109.7 & 58 $\pm$ 7 & 4 $\pm$ 2 \\ 
& 3- +100 \% & 1271.2 & 170.3 & 220.3 & 123.3 & 61.7 $\pm$ 7.5 & 1 $\pm$ 1.5 \\ 
& 4- +200 \% & 1278.7 & 177.5 & 221.0 & 131.3 & 62.3 $\pm$ 7.7 & 0.3 $\pm$ 1 \\ 
& 5- +500 \% & 1279.6 & 202.6 & 220.3 & 139.7 & 63.7 $\pm$ 7.9 & 0 $\pm$ 0.3 \\ \hline
\end{tabular}
\caption{Relaxed weekly rest constraint for moderate scenario}
\label{tab:relaxed_weekly}
\end{table}

We observe in Table \ref{tab:relaxed_weekly} that by relaxing the rest constraints on on-call duties, we allow for a better coverage of the demand (no shortage when $\alpha$ is large). More interestingly, the total effective work hours are reduced compared to schedules with tighter constraints. This corroborate our assertion that on-call duties can lead to a better coverage of the demand with less effective work hours being done.

\section{Conclusion}

There are very few recourse ED managers have to improve the quality of service during epidemic seasons. On-call duties are one of these possible recourse actions that are readily available. In EDs, on-calls are sometimes associated with extra work and unfriendly work conditions. We have analyzed in this paper the advantages and shortcoming of using on-call duties in an ED department for different epidemic scenarios. 

To this end, we developed a stochastic model to compare different scheduling strategies. The schedules obtained are robust in the sense that they hedge against the uncertainty in the epidemic realization if the type of the epidemic is known. Given the objective of minimizing the total costs incurred and covering the demand as best as possible, using on-call duties resulted in schedules that dominated the regular schedules.

We recommend using on-calls as a recourse mechanism during seasonal epidemic periods as they increase the coverage while keeping the costs down and prevent idle hours of employees. For instance, when resource shortage is very expensive, using on-call duties can lead to more than 30\% reduction in effective work hours with schedules that are 10\% less expensive. It is more desirable to use on-calls during periods with a lot of uncertainty on the demand. In the tests we ran this corresponded to day periods which had more variance. Nonetheless, in the scenarios considered, there is always over 60\% of regular duties in the optimal schedules which indicates that on-call duties are not a substitute to regular duties but rather an efficient mean to overcome increased variance in the workload.

Several extensions to this work are possible. The daily shift assignment problem  is necessary for determining the optimal allocation of shifts to resources within the day. Also, we did not consider in this study the interaction between the resources but non negligible queuing effects are present when determining daily shift assignments.  Furthermore, in this study we only focused on the workload resulting from epidemic patients, the workload patterns should be adjusted if different arrival trends are observed. For example, in some hospitals, the workload during the weekend is higher than during the week. This can easily be included when considering the demand scenarios to consider.

\section{acknowledgments}
This research is partially supported by the French National Research Agency ``ANR'' under grant number ANR-11-TECS-0010 of project HOST.

\section{Bibliography}


\end{document}